\documentclass[twocolumn,amsmath,showpacs,pra]{revtex4}
\usepackage{graphicx}

\newcommand{\degree}{\ensuremath{^{\circ}}}
\newcommand{\sqrthz}{\ensuremath{\sqrt{\textnormal{Hz}}}}
\newcommand{\GammaG}{\ensuremath{\Gamma_{\textnormal{G}}}}
\newcommand{\GammaL}{\ensuremath{\Gamma_{\textnormal{L}}}}
\begin{document}

\title{High-Temperature Alkali Vapor Cells with Anti-Relaxation Surface Coatings}

\author{S. J. Seltzer}
\altaffiliation{Present address: Department of Chemistry, University of California, Berkeley, CA 94720 and Materials Sciences Division, Lawrence Berkeley National Laboratory, Berkeley, CA 94720}

\author{M. V. Romalis}

\affiliation{Department of Physics, Princeton University, Princeton, New Jersey 08544}

\begin{abstract}
Anti-relaxation surface coatings allow long spin relaxation times in alkali-metal cells without buffer gas, enabling faster diffusion of the alkali atoms throughout the cell and giving larger signals due to narrower optical linewidths. Effective coatings were previously unavailable for operation at temperatures above 80\degree C. We demonstrate that octadecyltrichlorosilane (OTS) can allow potassium or rubidium atoms to experience hundreds of collisions with the cell surface before depolarizing, and that an OTS coating remains effective up to about 170\degree C for both potassium and rubidium. We consider the experimental concerns of operating without buffer gas and with minimal quenching gas at high vapor density, studying the stricter need for effective quenching of excited atoms and deriving the optical rotation signal shape for atoms with resolved hyperfine structure in the spin-temperature regime. As an example of a high-temperature application of anti-relaxation coated alkali vapor cells, we operate a spin-exchange relaxation-free (SERF) atomic magnetometer with sensitivity of 6~fT/\sqrthz\ and magnetic linewidth as narrow as 2~Hz. Copyright 2009 American Institute of Physics. This article may be downloaded for personal use only. Any other use requires prior permission of the author and the American Institute of Physics. The following article appeared in Journal of Applied Physics and may be found at http://link.aip.org/link/?jap/106/114905.
\end{abstract}


\maketitle

\section{Introduction}

Polarized alkali atoms lose their spin coherence after colliding with the walls of a glass vapor cell, limiting their spin-polarization lifetime and necessitating the use of either an anti-relaxation surface coating or a buffer gas in order to suppress the effects of wall relaxation. Paraffin is the most effective known coating for alkali vapor cells, allowing atoms to bounce up to 10,000 times off the cell wall without depolarizing \cite{Bouchiat1966}, and it is used in applications including magnetometry \cite{Budker2007}, frequency reference \cite{Budker2005}, spin squeezing \cite{Kuzmich2000}, slow light \cite{Klein2006}, and quantum memory \cite{Julsgaard2004}. In particular, modern alkali-metal magnetometers offer comparable or superior sensitivity to other magnetic field detectors, leading to recent interest in the study of the anti-relaxation properties of paraffin and other materials \cite{Zhao2008, Yi2008, Seltzer2008, Rampulla2009, Michalak2009}. Much of this effort has focused on finding another coating as effective as paraffin but with better high-temperature stability, since paraffin melts and is ineffective above roughly 80\degree C. The most sensitive atomic magnetometers operate at higher temperatures than this in order to achieve large vapor density (typically 10$^{12}$-10$^{14}$~cm$^{-3}$), including spin-exchange relaxation-free (SERF) vector magnetometers for detection of very weak fields \cite{Allred2002, Kominis2003}, scalar magnetometers for precise measurement of the amplitude of fields \cite{Smullin2006}, and radio-frequency magnetometers for detection of oscillating fields \cite{Savukov2005a, Lee2006}. Due to the unavailability of appropriate surface coatings, all previous implementations of these high-density alkali-metal magnetometers have instead used cells filled with buffer gas, which slows diffusion of polarized alkali atoms to the cell walls.

It was recently shown \cite{Seltzer2007} that a coating of octadecyltrichlorosilane (OTS) can allow potassium atoms to bounce off the cell wall up to 2,000 times without depolarizing, although the quality of the coating is highly variable and levels of several hundred bounces are more typical. The OTS molecule resembles paraffin, containing an 18-carbon alkane chain, except with a silicon head group that chemically binds to the glass surface, enabling an OTS film to survive intact at higher temperatures than a paraffin film. Here we demonstrate that a multilayer OTS coating operates for potassium and rubidium vapor at temperatures up to about 170\degree C, at which point we observe aging of the coating. We present the first use of anti-relaxation coated cells for high-density atomic magnetometry by implementing a SERF magnetometer using OTS-coated vapor cells. We discuss the advantages of coated cells over cells with buffer gas for high-density magnetometry, as well as experimental concerns such as more stringent quenching gas pressure requirements, and we derive the optical rotation signal profile in the spin-temperature regime for atoms with resolved hyperfine structure.

\section{Advantages of Coated Cells for Magnetometry and Practical Considerations}

Coated cells have several practical advantages over buffer gas cells that allow for easier realization of high magnetic field sensitivity. In the absence of polarization-destroying collisions with buffer gas atoms, the magnetic linewidth can be much narrower. Coatings are particularly advantageous for miniature cells, which otherwise require very high buffer gas pressure to prevent wall collisions, resulting in broad linewidths \cite{Kitching2002}. An anti-relaxation coating can be considered ``sufficient'' for a given alkali metal and cell size if it operates at high enough temperature that the spin-polarization lifetime is dominated by alkali-metal spin-destruction collisions at this temperature rather than wall collisions. For example, for a spherical cell with radius $r_0$, the number of bounces required to satisfy this condition is $N>3/[4 r_0 \sigma_{sd} n(T)]$, where $\sigma_{sd}$ is the alkali-metal spin destruction cross-section and $n(T)$ is the density of alkali-metal atoms as a function of temperature. Once the vapor density is high enough to meet this condition, any improvement in the coating quality will not affect significantly the lifetime. Under these conditions the magnetometer can reach the fundamental sensitivity limit set by the spin-destruction cross-section of the alkali-metal atoms. Here for the first time we demonstrate a coating of sufficient quality for Rb atoms in a $\sim$1~cm cell.

In a coated cell containing only a moderate amount (10-50~Torr) of quenching gas to prevent radiation trapping, the individual alkali atoms diffuse quickly enough that they sample a large fraction of the cell volume during each polarization lifetime. The entire cell may thus be used as the active measurement volume without needing to expand the pump beam to fill the cell, since the wall coating allows some degree of polarization to be maintained even in the parts of the cell not actively being pumped. The probe beam likewise does not need to be greatly expanded because a large number of atoms are likely to pass through it during a polarization lifetime. Higher coating efficiency permits the use of smaller pump and probe beams. In contrast, the atoms diffuse very slowly in cells containing high buffer gas pressure and therefore are localized in small regions of the cell during each polarization lifetime, so that the measurement volume consists only of the intersection of the pump and probe beams. Rapid diffusion of atoms throughout a coated cell also enables partial suppression of magnetic field gradient effects on the polarization lifetime through motional averaging \cite{Cates1988}.

In addition, the lack of pressure broadening due to buffer gas results in narrower optical resonance linewidths in coated cells. The integral of the photon absorption cross-section $\sigma (\nu )$ associated with a given resonance is constant regardless of the lineshape,
\begin{equation}
\int^{+\infty}_{0}\sigma (\nu )\,\textnormal{d}\nu =\pi r_{e}cf, \label{eq_int_constant}
\end{equation}
where $r_{e}$ is the classical electron radius, and $f$ is the oscillator strength of the transition (see for example \cite{Corney1977}). A narrower resonance linewidth therefore results in stronger interaction between the atoms and the pump and probe beams. The combined effect of Doppler broadening, which has a Gaussian profile with width (FWHM) \GammaG, and the pressure broadening due to the buffer/quenching gas, which has a Lorentzian profile with width \GammaL, is described by a Voigt profile, as discussed in the Appendix. If the ground- and excited-state hyperfine structure of the alkali atom is partially or fully resolved, as is generally the case for cells with minimal pressure broadening, then it is necessary to consider separately the transitions between individual energy levels of the ground and excited states.

\begin{figure}
\centering
\includegraphics[width=8cm]{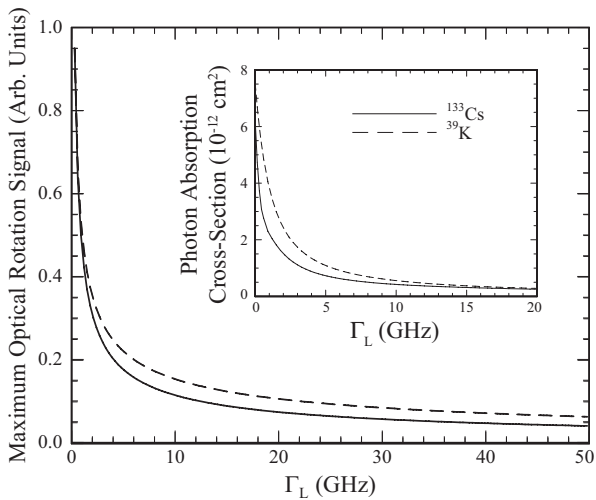}
\caption{Scaling of the maximum optical rotation signal $\theta (\nu )e^{-n\sigma(\nu )l}$ of the probe beam, taken at optimal detuning from resonance, as a function of buffer gas pressure broadening (FWHM) \GammaL . The potassium and cesium curves are normalized separately, and \GammaG\ is set to its value at 160\degree C. The inset shows the maximum photon absorption cross-section as a function of \GammaL .} \label{fig_signalcomp}
\end{figure}

Coated cells require significantly less laser power than buffer gas cells to achieve the same optical pumping rate, as a result of decreased pressure broadening of the optical linewidth. Wall coatings are therefore highly favorable for portable and miniaturized devices where power consumption must be minimized \cite{Knappe2006}. The inset of Figure~\ref{fig_signalcomp} shows the maximum absorption cross-sections of $D1$ light calculated from Equation~\ref{eq_sigma} as a function of \GammaL\ for $^{133}$Cs and $^{39}$K atoms at 160\degree C (\GammaG =0.9~GHz for $^{39}$K and 0.4~GHz for $^{133}$Cs). For narrow optical linewidths such that the hyperfine structure of the atom is partially resolved, it is necessary to determine from Equation~\ref{eq_sigma} the frequency of light that gives the largest cross-section for a particular linewidth \GammaL . The pressure broadening of the $D1$ resonance for potassium is 13.2~GHz/amg due to helium and 21.0~GHz/amg due to nitrogen \cite{Lwin1978}, while the broadening for cesium is 19.9~GHz/amg due to helium and 14.8~GHz/amg due to nitrogen \cite{Andalkar2002} (1~amg=1~atm at 0\degree C). The need for some small amount of nitrogen (about 10-50 Torr) for quenching in coated cells requires a minimum $\GammaL \lesssim$ 1~GHz, whereas high-density magnetometers typically employ up to several amg of helium as a buffer gas.

One of the primary challenges in achieving high magnetometer sensitivity is overcoming probe beam noise. The optical rotation of an off-resonant, linearly polarized laser beam can be used to measure the spin projection $P_{x}$ of the alkali atoms, as shown in Equation~\ref{eq_theta}, with the narrower optical linewidths of coated cells resulting in larger optical rotations. Noise in the measurement of beam polarization can obscure the rotation signal, so the use of coated cells can allow for easier detection of small alkali polarization signals amidst the background of probe beam noise. For example, at frequencies below about 100~Hz, convection of the air that the probe beam passes through causes motion of the beam. The resulting 1/f noise can impair magnetometer performance by more than an order of magnitude; methods to deal with this noise, including the use of evacuated beam paths \cite{Smullin2006} and modulation of the magnetic field to shift the magnetometer signal to a higher frequency \cite{Li2006}, complicate the operation of the magnetometer. Narrower optical lines of cells with little buffer gas also result in higher optical depth on resonance for a given cell length and alkali metal density. Hence they are more advantageous for quantum-non-demolition polarization measurements where the improvement in signal-to-noise ratio scales as the square root of the optical density on resonance.

A narrow optical linewidth also results in strong attenuation of the probe beam, so while very large optical rotations can be obtained by tuning the probe beam close to resonance, the resulting polarimeter signal, which is proportional to $\theta (\nu )e^{-n\sigma(\nu )l}$, may become small. Here $n$ is the density of the alkali vapor, and $l$ is the path length of the probe beam through the cell. The optimal probe beam detuning that maximizes this signal depends on the parameters of the system, including the buffer/quenching gas pressure and alkali density; if the hyperfine structure is optically resolved, then the individual resonances must be considered as in Equation~\ref{eq_theta}. Figure~\ref{fig_signalcomp} shows the scaling of the maximum polarimeter signal for $^{133}$Cs and $^{39}$K atoms as a function of \GammaL , for the case of a cell at 160\degree C and atomic density 5x10$^{12}$~cm$^{-3}$ with length 5~cm and $P=0.5$. The curves for the two isotopes are normalized separately and should not be directly compared. Coated cells show an increase in signal of up to an order of magnitude over cells with several amg of buffer gas.

There are however practical considerations that must be addressed when operating at high alkali vapor density using coated cells, due to the narrow optical linewidths. One of these considerations is that the vapor may be too optically thick to allow the pump beam to penetrate far into the cell. In this case, the wall coating must be sufficiently good and the atomic diffusion must be sufficiently fast that the atoms are able to sample the front of the cell and get pumped during each polarization lifetime. Otherwise, the average polarization throughout the cell can be increased by modulating the frequency of the pump beam, thus enabling the beam to travel further into the cell without significant attenuation during the time when it is modulated off resonance and the absorption rate is decreased. If the modulation is symmetric about the resonance, then there is no average light shift. The modulation amplitude should be chosen such that even the off-resonant light maintains a non-zero instantaneous pumping rate, in order that the light which penetrates to the back of the cell is able to pump the atoms there, and the modulation rate should be fast compared to the relaxation rate of the atoms to achieve significant polarization throughout the cell.

Another consideration is the appropriate amount of quenching gas to include in the cell because radiation trapping is more likely to limit the polarization in cells with a narrower optical linewidth. Buffer gas cells can generally contain more quenching gas than is strictly necessary, since high buffer gas pressure broadens the optical linewidth and limits the atomic polarization lifetime to a much greater degree than the relatively small quenching gas pressure. On the other hand, coated cells should not contain an excessive amount of quenching gas, so as not to unnecessarily slow atomic diffusion, broaden the optical resonances, or cause additional spin destruction.

The polarization $P$ attainable at low quenching gas pressure can be understood qualitatively from a modified version of the model given in \cite{Rosenberry2007}:
\begin{equation}
P=\frac{R_{\textnormal{OP}}}{R_{\textnormal{OP}}+R_{\textnormal{SD}}+K(M-1)QR_{\textnormal{OP}}(1-P)}\ ,\label{eq_trapping}
\end{equation}
where $R_{\textnormal{OP}}$ is the optical pumping rate, $R_{\textnormal{SD}}$ is the spin-destruction rate (including collisions with the wall, other alkali atoms, and quenching gas molecules), and $M$ is the number of times a photon is emitted before escaping the cell. $K\sim0.1$ is the degree of depolarization caused by a reabsorbed photon, which is determined by the nuclear slowing-down factor, as well as a geometrical factor due to some fraction of reemitted photons having the same polarization as the pumping photons. $Q$ is the probability of an alkali atom in the excited state decaying to the ground state by spontaneous decay rather than quenching, given by
\begin{equation}
Q=\frac{1}{1+R_{\textnormal{Q}} \tau}=\frac{1}{1+p_{\textnormal{Q}} /p_{\textnormal{Q}}'}\ .\label{eq_quenching}
\end{equation}
Here, $R_{\textnormal{Q}}$ is the rate of collisions with quenching gas molecules, $\tau$ is the excited-state lifetime, $p_{\textnormal{Q}}$ is the quenching gas pressure, and $p_{\textnormal{Q}}'$ is the characteristic pressure that gives $Q$=0.5. At 150\degree C the characteristic pressure of nitrogen for the $D1$ resonance is 6.3~Torr for potassium, 4.2~Torr for rubidium, and 3.7~Torr for cesium, as determined from the measured quenching cross-sections \cite{McGillis1967,McGillis1968,Hrycyshyn1970}. Less quenching gas is therefore required for operation with rubidium or cesium compared to potassium, perhaps making these elements better practical choices for some applications of high-density magnetometry using coated cells; however, potassium allows for better intrinsic sensitivity \cite{Allred2002}, so future study of quenching gas pressure requirements is called for.

In Equation~\ref{eq_trapping}, the last term in the denominator is the spin-relaxation rate due to absorption of reemitted photons, which we modify from the form given in \cite{Rosenberry2007} to include a factor of $(1-P)$ because polarized atoms can not absorb pump photons. This modification is necessary to account for the large polarization achieved in some applications at high alkali density; for example, radio-frequency magnetometers operate with nearly 100\% polarization at high density in order to suppress spin-exchange broadening \cite{Savukov2005a}. The modified expression for $P$ also better fits the high-density data in \cite{Rosenberry2007} than the original model.

\section{Experiment and Observations}

To demonstrate the effectiveness of coated cells in high-density magnetometry, we implemented a SERF magnetometer similar to those described previously \cite{Allred2002, Kominis2003}. Orthogonal pump and probe beams were generated by distributed feedback diode lasers that were easily tunable over a large frequency range by adjusting the temperature and/or current of the diode. The frequency of the pump beam was modulated as necessary to achieve optimal polarization throughout the cell. The double-walled hot air oven was surrounded by five layers of $\mu$-metal shields with a shielding factor of about 10$^{5}$, and magnetic field coils were used to cancel any residual fields.

The procedure for coating Pyrex glass with OTS was adapted from \cite{Rosen1999, Bear2000} and was described in detail in \cite{Seltzer2008}. The coatings were not prepared in the moisture-free conditions necessary for formation of a monolayer, and the thickness of the films has been measured to be roughly twice the length of an OTS molecule, indicating the presence of multiple layers at the surface \cite{Seltzer2008}. The coated potassium cells were comprised of 5~cm diameter spheres with 1~mm inner-diameter stems containing potassium metal in natural abundance; the opening of the stem into the cell was made as small as practical since polarized atoms are lost when they enter the stem.

In addition to the evacuated cells described in \cite{Seltzer2007}, we also made several additional potassium cells with either 5.0 or 12.6~Torr of nitrogen for quenching of excited atoms. We measured the quality of the coating in a particular cell by first polarizing the alkali atoms and then turning off the optical pumping and observing the subsequent polarization lifetime $T_{1}$. In order to determine the number of bounces $N$ allowed in a spherical cell of radius $r_{0}$ containing quenching gas, we consider diffusion with the boundary condition given by \cite{Masnou1967} for the coated surface:
\begin{equation}
\frac{\partial}{\partial r}P=\frac{-\overline{v}P}{2N(2-1/N)D}\ ,\label{eq_boundary}
\end{equation}
where $P$ is the polarization as a function of position and time, $\overline{v}$ is the thermal velocity of the alkali atoms, and $D$ is the diffusion constant. The solution to the diffusion equation for the fundamental diffusion mode is given by the following system of equations, which may be solved numerically:
\begin{eqnarray}
k\cot(k\hspace{1.5pt}r_{0})&=&\frac{1}{r_{0}}-\frac{\overline{v}}{2N(2-1/N)D}\ ,\label{eq_solutionA} \\
\frac{1}{T_{1}}&=&D\hspace{1.5pt}k^{2}+R_{\textnormal{col}}\hspace{1.0pt},\label{eq_solutionB}
\end{eqnarray}
where $k$ is the radial diffusion wave number, and $R_{\textnormal{col}}$ is the spin-destruction rate due only to alkali-alkali and alkali-quenching gas collisions. A similar solution may be obtained for arbitrary cell geometry. The new cells exhibited relaxation times in the range of 30-45~ms, consistent with 400-600~bounces, though none matched the 2,000~bounces allowed by one of the evacuated cells. We also measured the polarization lifetimes in two OTS-coated rubidium cells to both be consistent with 700-900~bounces. Recent observations of both monolayer and multilayer OTS suggest that the quality of the film as an anti-relaxation coating in a particular cell may be determined by the fractional coverage of defect sites on the surface \cite{Rampulla2009}. The estimated numbers of bounces are only approximate due to the uncertainty in the values of the diffusion constants of potassium and rubidium in nitrogen; we use $D$=0.25~cm$^{2}$/s for potassium \cite{Silver1984} and $D$=0.14~cm$^{2}$/s for rubidium \cite{Franz1976} (at 273~K). The OTS-coated cells made in our lab for this study and for related studies are listed in Table~\ref{table_cells}.

\begin{table*}
\begin{center}
{\addtolength{\tabcolsep}{5 pt}\scriptsize
\begin{tabular}{cccccccc}
\hline\hline
Cell & Size and Shape & Alkali & Gas & Max. Polarization & Max. T$_{1}$ (ms) & Bounces (Approx.)  \tabularnewline
\hline
1 & 0.90'' Diameter Sphere & K & 26.0 Torr N$_{2}$ & & 23 & 500  \tabularnewline
2 & 0.90'' Diameter Sphere & K & 26.0 Torr N$_{2}$ & & 24 & 550  \tabularnewline
4 & 1.90'' Diameter Sphere & K & None & & 1.6 & 20  \tabularnewline
5 & 1.90'' Diameter Sphere & K & None & 2\% & 33 & 500 \tabularnewline
6 & 1.95'' Diameter Sphere & K & None & & 61 & 900 \tabularnewline
7 & 1.96'' Diameter Sphere & K & None & & 145 & 2100 \tabularnewline
8 & 2.00'' Diameter Sphere & K & 5.0 Torr N$_{2}$ & & 40 & 500 \tabularnewline
9 & 2.00'' Diameter Sphere & K & 5.0 Torr N$_{2}$ & & 33 & 400 \tabularnewline
10 & 2.01'' Diameter Sphere & K & 5.0 Torr N$_{2}$ & 25\% & 46 & 600 \tabularnewline
11 & 2.00'' Diameter Sphere & K & 12.6 Torr N$_{2}$ & 45\% & 45 & 450 \tabularnewline
12 & 2.2 cm $\times$ 2.2 cm $\times$ 3.5 cm & $^{87}$Rb & 5 Torr N$_{2}$ & & 51 & 850 \tabularnewline
13 & 2.5 cm $\times$ 2.5 cm $\times$ 4.5 cm & $^{87}$Rb & 10 Torr N$_{2}$ & 80\% & 59 & 700 \tabularnewline
\hline\hline
\end{tabular}
}
\end{center}
\caption{List of OTS-coated cells produced for this study and for related studies. The polarization is given for maximum pump beam intensity at a vapor density of 5$\times$10$^{12}$~cm$^{-3}$, and the polarization lifetime $T_{1}$ is given as the maximum value measured for each particular cell.}
\label{table_cells}
\end{table*}

We found that the OTS coating could withstand temperatures below 160\degree C without permanent damage in the presence of both potassium and rubidium vapor, although cells operating at 170\degree C displayed slow degradation of coating performance on a time scale of days. We measured the alkali density to be less than expected from the saturated vapor pressure and found that it varied by as much as a factor of 2 from cell to cell, with observed densities at 170\degree C up to 9$\times$10$^{12}$~cm$^{-3}$ for potassium and 9$\times$10$^{13}$~cm$^{-3}$ for rubidium. The maximum operating temperature is the same for both species despite the difference in density by up to an order of magnitude, suggesting that the coating is damaged by collisions between individual alkali atoms and the silane head groups attached to the surface; some minimum thermal energy may be required to break the chemical bond between the OTS molecules and the glass. For comparison, a monolayer OTS coating remains undamaged up to 190\degree C in the presence of rubidium \cite{Yi2008}, and it can survive temperatures up to 400\degree C when not exposed to alkali vapor \cite{Fedchak1997}.

\begin{figure}
\centering
\includegraphics[width=8cm]{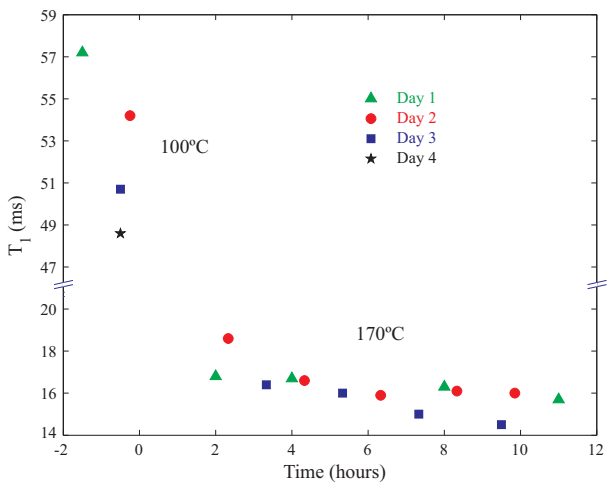}
\caption{(Color online) Measurements of the polarization lifetime $T_{1}$ taken in an OTS-coated rubidium cell (\#13) with 10~Torr of nitrogen quenching gas, showing degradation of the coating after extended operation at 170\degree C. Note that at 170\degree C the relaxation time is dominated by collisions between rubidium atoms, indicating that the surface coating is sufficiently effective.} \label{fig_decaytime}
\end{figure}

Figure~\ref{fig_decaytime} shows measurements of the polarization lifetime $T_{1}$ over the course of several days in a rubidium cell. The temperature was increased from 100\degree C to 170\degree C at $t$=0, and it was returned to 100\degree C after the final measurement was taken on each day. When at lower temperature, the stem was actively cooled in order to collect the alkali metal; this was found to reverse temporary reductions of the polarization lifetime due to alkali condensation on the walls of the main cell body. At higher temperature $T_{1}$ is dominated by spin-destruction collisions between rubidium atoms and is thus highly sensitive to fluctuations in the vapor density, but a permanent reduction in the coating effectiveness after each day of operation at 170\degree C is evident from comparing the data taken at lower temperature. The polarization lifetime never recovered, even after leaving the cell at lower temperature with its stem cooled for a period of several days. We speculate that patches of bare glass are slowly exposed as OTS molecules are removed from the surface, resulting in gradual degradation of the coating. This is the first time that a sufficient anti-relaxation coating has been demonstrated with rubidium vapor in a cell of this size, and it raises the possibility of a rubidium SERF magnetometer with an OTS-coated cell reaching its fundamental sensitivity limit. A coating would need to operate at higher temperature to be sufficient for potassium vapor, although the use of potassium would allow for an intrinsically more sensitive magnetometer because of its smaller spin-destruction cross-section.

\begin{figure}
\centering
\includegraphics[width=8cm]{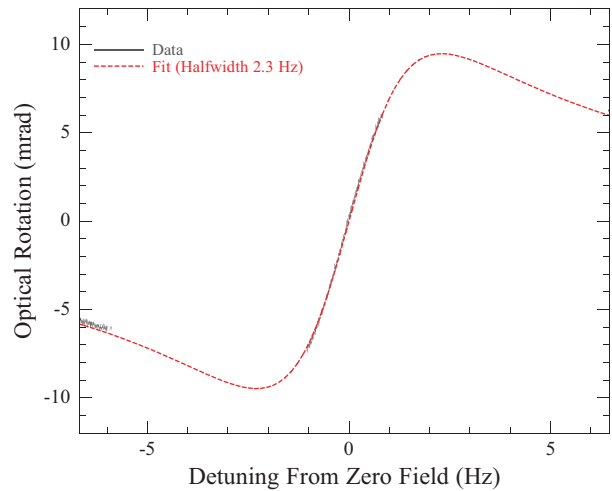}
\caption{(Color online) A narrow magnetometer resonance signal obtained in the SERF regime with an OTS-coated cell (\#7) and very low optical pumping rate, fitted to a Lorentzian with HWHM of 2.3~Hz.} \label{fig_narrowline}
\end{figure}

The long relaxation time allowed by an OTS coating enables magnetic resonance linewidths comparable to those observed in low-temperature paraffin cells \cite{Budker2005} and in buffer gas cells in the SERF regime \cite{Allred2002}. Spin-exchange relaxation is eliminated in a SERF magnetometer by operating at high vapor density \cite{Happer1977}, so the magnetic linewidth $\Delta\nu$ is given by the optical pumping and spin-destruction rates as
\begin{equation}
2\pi\times\Delta\nu=1/T_{2}=(R_{\textnormal{OP}}+R_{\textnormal{SD}})/q,\label{eq_deltanu}
\end{equation}
where the linewidth is given as the halfwidth at half-maximum (HWHM), and $q$ is the polarization-dependent nuclear slowing-down factor (see \cite{Appelt1998}). Near zero field there is no well-defined quantization axis, so $T_{1}\cong T_{2}$, and the measured value of $T_{1}$ in a particular cell gives the narrowest magnetic linewidth that can be attained in the cell, although in practice the linewidth tends to be larger than this minimum value. Figure~\ref{fig_narrowline} shows a resonance signal with linewidth of 2.3~Hz obtained at 150\degree C using the evacuated potassium cell that allows 2,000~bounces (Cell \#7), taken with $R_{\textnormal{OP}}\ll R_{\textnormal{SD}}$ so that the linewidth is dominated by the collisions with the OTS-coated surface. Although radiation trapping limited the alkali polarization in this cell to about 2$\%$ even with high pumping rate, similar performance should be possible in cells with sufficient quenching gas and comparable coating effectiveness.

\begin{figure}
\centering
\includegraphics[width=8cm]{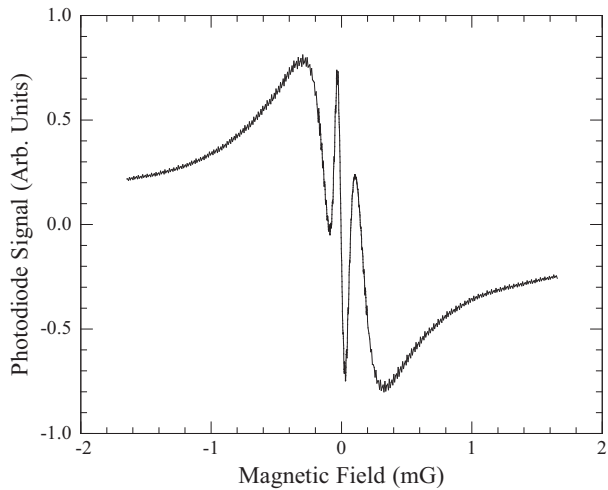}
\caption{A dispersion curve with amplitude of 1.6~radians measured in a coated cell (\#11); rotations larger than $\pi$/4 radians cause the polarimeter signal to turn over.} \label{fig_largerotation}
\end{figure}

Radiation trapping also limited the polarization in the cells with quenching gas. The pumping rate should be equal to the spin-destruction rate for optimal performance of a SERF magnetometer, giving 50\% polarization, but under this condition the polarization in the potassium cells was observed to be only about 5-10$\%$ at a density of $n$=5$\times$10$^{12}$~cm$^{-3}$. In contrast, in a rubidium cell with 10~Torr of nitrogen at the same density, we observe the expected polarization. Even with such low polarization, large optical rotation signals were observed in the potassium cells. A dispersion curve with an amplitude of 1.6~radians is shown in Figure \ref{fig_largerotation}, exhibiting the characteristic turn-over of the polarimeter signal once the optical rotation angle reaches $\pi$/4. Such large rotations have not been seen previously in SERF magnetometers using buffer gas-filled cells.

Magnetometers often operate using a feedback loop to lock to the zero crossing of the resonant dispersion curve, so a useful figure-of-merit is the slope at this point. A SERF magnetometer with its pump beam in the $\hat{z}$ direction and its probe beam in the $\hat{x}$ direction is primarily sensitive to the magnetic field component along $\hat{y}$; if this field is sufficiently small that the polarization remains nearly along $\hat{z}$, then the magnetometer signal $\theta$ given by Equation~\ref{eq_theta} is proportional to
\begin{equation}
P_{x}\approx P_{z}\frac{\gamma^{e}B_{y}(R_{\textnormal{OP}}+R_{\textnormal{SD}})}{(R_{\textnormal{OP}}+R_{\textnormal{SD}})^{2}+(\gamma^{e}B_{y})^{2}}\ ,\label{eq_Px}
\end{equation}
where $\gamma^{e}=g_{s}\mu_{B}/\hbar$ is the gyromagnetic ratio of the electron \cite{Allred2002}. In a potassium cell with 12.6~Torr of nitrogen and pumping rate equal to the spin-destruction rate, giving 8$\%$ polarization and HWHM of 6.5~Hz at $n$=5$\times$10$^{12}$ cm$^{-3}$, we measured a slope $\partial\theta/\partial B_{y}$ of 57~rad/mG near the point of optimal probe beam detuning, in agreement with Equations~\ref{eq_Px} and~\ref{eq_theta}. Tuning the probe beam closer to the optical resonance, we achieved a slope of 135~rad/mG. For comparison, in a square cell with length 2.2~cm and 2.9~amg of helium buffer gas at the same density, 50$\%$ polarization, and HWHM of 4.6~Hz, we measured a slope of 15~rad/mG at the point of optimal detuning, close to the predicted 17~rad/mG. Even accounting for the longer optical path length, the slope was larger in the coated cell despite a broader magnetic linewidth and significantly smaller polarization. The advantage should be even greater in cells with additional nitrogen to attain higher polarization while still maintaining a narrow linewidth \GammaL .

\begin{figure}
\centering
\includegraphics[width=8cm]{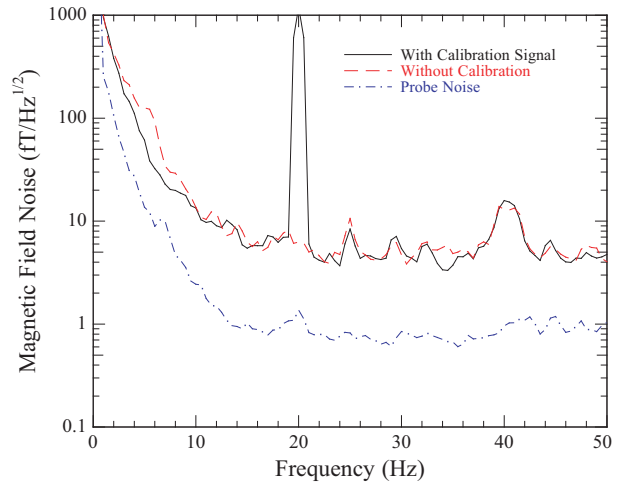}
\caption{(Color online) Single-channel noise spectrum taken at 150\degree C in a coated potassium cell without quenching gas, showing sensitivity of 6~fT/\sqrthz , possibly limited by the noise of the magnetic shielding. The optical noise of the probe beam was measured in the absence of pumping light, so that the vapor was unresponsive to ambient field noise.} \label{fig_spectrum}
\end{figure}

Shown in Figure~\ref{fig_spectrum} is a magnetic field noise spectrum taken in an OTS-coated potassium cell without buffer gas at 150\degree C. The magnetometer sensitivity is 6~fT/\sqrthz , possibly limited by the noise of the magnetic shielding, even with the low polarization attained in the cell. We achieved sensitivity better than 10~fT/\sqrthz\ at 120\degree C, a significantly lower temperature than previously used in a potassium SERF magnetometer. Magnetic field sensitivity can be improved by up to an order of magnitude by measuring the field gradient, thus canceling out common-mode noise between different channels. This is easily accomplished in a cell where buffer gas limits the diffusion of alkali atoms, permitting different parts of the cell to act as individual measurement volumes \cite{Kominis2003}. It should be possible to make a similar multiple-channel gradient measurement using a coated cell containing separate chambers.

\section{Conclusion}

In summary, we have found that a multilayer OTS coating preserves alkali polarization at temperatures up to 170\degree C for both potassium and rubidium vapor, despite an order of magnitude difference in vapor density at that temperature, motivating further investigation of the interaction between alkali atoms and the OTS surface. We have demonstrated the use of coated cells in high-density alkali-metal magnetometry, showing that such cells permit narrow magnetic resonance linewidths and feature larger optical rotation signals than buffer gas-filled cells under similar operating conditions. An OTS coating that allows several hundred bounces was found to be sufficient for rubidium cells with dimensions of several~cm near the maximum operating temperature.

We have also studied some of the experimental concerns specific to high-density alkali vapor with minimal pressure broadening, including the stricter requirements for quenching that cause additional depolarization from radiation trapping; indeed, radiation trapping limited the polarization in the potassium cells used for this study, so cells with additional quenching gas should show an even larger improvement over cells with buffer gas. All measurements were made with a SERF magnetometer, but the results are equally valid for radio-frequency and scalar magnetometers operating at high density. In the Appendix we derive the shape of the optical rotation signal for alkali atoms with resolved hyperfine lines in the spin-temperature regime, which is typically the case for atoms at high density without pressure broadening from buffer gas. Anti-relaxation coatings will be an essential tool as the sensitivity of high-density magnetometers drives toward the attotesla level, underscoring the importance of developing more effective and reliable high-temperature surface coatings.

The authors acknowledge Tom Kornack for coating the rubidium cells. This work was funded by an Office of Naval Research MURI grant.

\appendix*
\section{Optical rotation signals with resolved hyperfine structure in the spin-temperature regime}

\begin{table*}
\begin{center}
{\addtolength{\tabcolsep}{3 pt}
\begin{tabular}{c|cc|cc|cc}
\hline\hline
 & \multicolumn{2}{c}{$I=3/2$} & \multicolumn{2}{c}{$I=5/2$} & \multicolumn{2}{c}{$I=7/2$}  \tabularnewline
\raisebox{1.5ex}[0pt]{Transition} & $A_{F_{g}F_{e}}$\rule[-1.2ex]{0pt}{0pt} & $B_{F_{g}F_{e}}$ & $A_{F_{g}F_{e}}$ & $B_{F_{g}F_{e}}$ & $A_{F_{g}F_{e}}$ & $B_{F_{g}F_{e}}$ \tabularnewline
\hline
$I-\frac{1}{2}\rightarrow I-\frac{1}{2}$ & 1/16 & \rule{0pt}{3.6ex}\rule[-1.2ex]{0pt}{0pt} $\frac{1-P^{2}}{16(1+P^{2})}$ & 5/54 & $\frac{5-2P^{2}-3P^{4}}{18(3+10P^{2}+3P^{4})}$ & 7/64 & $\frac{7+7P^{2}-11P^{4}-3P^{6}}{64(1+7P^{2}+7P^{4}+P^{6})}$ \tabularnewline
$I-\frac{1}{2}\rightarrow I+\frac{1}{2}$ & 5/16 & \rule{0pt}{3.6ex}\rule[-1.2ex]{0pt}{0pt} $\frac{-5(1-P^{2})}{16(1+P^{2})}$ & 35/108 & $\frac{-7(5-2P^{2}-3P^{4})}{18(3+10P^{2}+3P^{4})}$ & 21/64 & $\frac{-9(7+7P^{2}-11P^{4}-3P^{6})}{64(1+7P^{2}+7P^{4}+P^{6})}$ \tabularnewline
$I+\frac{1}{2}\rightarrow I-\frac{1}{2}$ & 5/16 & \rule{0pt}{3.6ex}\rule[-1.2ex]{0pt}{0pt} $\frac{3(5+3P^{2})}{16(1+P^{2})}$ & 35/108 & $\frac{10(7+14P^{2}+3P^{4})}{18(3+10P^{2}+3P^{4})}$ & 21/64 & $\frac{7(15+63P^{2}+45P^{4}+5P^{6})}{64(1+7P^{2}+7P^{4}+P^{6})}$ \tabularnewline
$I+\frac{1}{2}\rightarrow I+\frac{1}{2}$ & 5/16 & \rule{0pt}{3.6ex}\rule[-2.2ex]{0pt}{0pt} $\frac{5+3P^{2}}{16(1+P^{2})}$ & 7/27 & $\frac{2(7+14P^{2}+3P^{4})}{18(3+10P^{2}+3P^{4})}$ & 15/64 & $\frac{15+63P^{2}+45P^{4}+5P^{6}}{64(1+7P^{2}+7P^{4}+P^{6})}$ \tabularnewline
\hline\hline
\end{tabular}
}
\end{center}
\caption{Relative strengths of the individual $D1$ transitions $F_{g}\rightarrow F_{e}$ for photon absorption ($A_{F_{g}F_{e}}$) and optical rotation ($B_{F_{g}F_{e}}$).}
\label{table_coeff}
\end{table*}

In cells with no buffer gas and a limited amount of quenching gas, the pressure broadened linewidth \GammaL\ is comparable to the Doppler broadened linewidth \GammaG. For light of frequency $\nu$, the optical resonance lineshape is then described by a Voigt profile,
\begin{equation}
V(\nu -\nu_{0})=\frac{2\sqrt{\ln{2}/\pi}}{\GammaG}w(\frac{2\sqrt{\ln{2}}[(\nu -\nu_{0}) +i\GammaL /2]}{\GammaG}), \label{eq_Voigt}
\end{equation}
\begin{equation}
w(x)=e^{-x^{2}}(1-\textnormal{erf}(-ix)), \label{eq_erf}
\end{equation}
where \GammaL\ and \GammaG\ are both given as full-width at half maximum (FWHM), $\nu_{0}$ is the resonance frequency, and $w(x)$ is the complex error function. For light near the $D1$ transition of an atom with nuclear spin $I$, the ground- and excited-state hyperfine levels are $F_{g}=I\pm 1/2$ and $F_{e}=I\pm 1/2$, respectively, and the resonance frequency for the transition $F_{g}\rightarrow F_{e}$ is $\nu_{F_{g}F_{e}}$. The weight of this transition for unpolarized atoms is
\begin{equation}
A_{F_{g}F_{e}}=\frac{(2F_{g}+1)(2F_{e}+1)}{2I+1} \left\{\begin{array}{ccc} 1/2 & 1 & 1/2 \\ F_{e} & I & F_{g} \end{array}\right\}^{2}, \label{eq_weight}
\end{equation}
where the curly brackets denote the 6-$j$ symbol. The coefficients $A_{F_{g}F_{e}}$ are listed in Table~\ref{table_coeff}. The photon absorption cross-section $\sigma (\nu )$ is given by \cite{Happer1967}
\begin{equation}
\sigma (\nu )=\pi r_{e}cf \sum_{F_{g},F_{e}}A_{F_{g}F_{e}}\ \textnormal{Re}[V(\nu -\nu_{F_{g}F_{e}})], \label{eq_sigma}
\end{equation}
where $r_{e}$ is the classical electron radius, and $f$=1/3 is the oscillator strength.

Similarly, the optical rotation angle $\theta (\nu )$ of a linearly polarized probe beam propagating in the $\hat{x}$-direction is determined by the difference in the absorption cross-sections of $\sigma^{+}$ and $\sigma^{-}$ light. For atoms in the spin-temperature distribution \cite{Anderson1960}, the relative weight of the transition $F_{g}\rightarrow F_{e}$ is therefore
\begin{eqnarray}
B_{F_{g}F_{e}}&\propto& A_{F_{g}F_{e}}\hspace{-2.0pt} \sum_{m_{F}}e^{\beta m_{F}}\left[\left( \begin{array}{ccc} F_{e} & 1 & F_{g} \\ m_{F}+1 & -1 & -m_{F} \end{array}\right) ^{2}\right.\nonumber\\
&&\hspace{40.0pt}-\left.\left( \begin{array}{ccc} F_{e} & 1 & F_{g} \\ m_{F}-1 & +1 & -m_{F} \end{array}\right)^{2} \hspace{2.0pt}\right] ,\label{eq_rotation}
\end{eqnarray}
where the sum is over all Zeeman sublevels of the ground state $F_{g}$, and $\beta=\ln[(1+P)/(1-P)]$ is the spin-temperature parameter. These coefficients are themselves functions of the alkali spin polarization, and they are listed in Table~\ref{table_coeff} in normalized form such that $\sum_{F{g},F_{e}}B_{F_{g}F_{e}}=1$. The total rotation angle is then given by
\begin{equation}
\theta (\nu )=\frac{\pi}{2}lr_{e}cfnP_{x} \sum_{F_{g},F_{e}}B_{F_{g}F_{e}}\ \textnormal{Im}[V(\nu -\nu_{F_{g}F_{e}})], \label{eq_theta}
\end{equation}
where $l$ is the path length of the probe beam through the cell, $n$ is the density of alkali atoms, and $P_{x}=\vec{P}\cdot \hat{x}$ is the component of alkali polarization along the probe beam direction. Equations~\ref{eq_sigma} and~\ref{eq_theta} may also be used to describe the case of large pressure broadening due to high pressures of buffer gas, though $V(\nu -\nu_{0})$ becomes nearly Lorentzian when $\GammaL \gg\GammaG$.


\begin{thebibliography}{}

\bibitem{Bouchiat1966} M. A. Bouchiat and J. Brossel, Phys. Rev. \textbf{147}, 41 (1966).

\bibitem{Budker2007} D. Budker and M. Romalis, Nature Phys. \textbf{3}, 227 (2007).


\bibitem{Budker2005} D. Budker, L. Hollberg, D. F. Kimball, J. Kitching, S. Pustelny, and V. V. Yashchuk, Phys. Rev. A \textbf{71}, 012903 (2005).



\bibitem{Kuzmich2000} A. Kuzmich, L. Mandel, and N. P. Bigelow, Phys. Rev. Lett. \textbf{85}, 1594 (2000).

\bibitem{Klein2006} M. Klein, I. Novikova, D. F. Phillips, and R. L. Walsworth, J. Mod. Opt. \textbf{53}, 2583 (2006).

\bibitem{Julsgaard2004} B. Julsgaard, J. Sherson, J. I. Cirac, J. Fiur\'{a}\u{s}ek, and E. S. Polzik, Nature \textbf{432}, 482 (2004).


\bibitem{Zhao2008} K. F. Zhao, M. Schaden, and Z. Wu, Phys. Rev. A \textbf{78}, 034901 (2008).

\bibitem{Yi2008} Y. W. Yi \textit{et al.}, J. Appl. Phys. \textbf{104}, 023534 (2008).

\bibitem{Seltzer2008} S. J. Seltzer, D. M. Rampulla, S. Rivillon-Amy, Y. J. Chabal, S. L. Bernasek, and M. V. Romalis, J. Appl. Phys. \textbf{104}, 103116 (2008).


\bibitem{Rampulla2009} D. M. Rampulla, N. Oncel, E. Abelev, Y. W. Yi, S. Knappe, and S. L. Bernasek, Appl. Phys. Lett. \textbf{94}, 041116 (2009).

\bibitem{Michalak2009} D. J. Michalak \textit{et al.}, in preparation.

\bibitem{Allred2002} J. C. Allred, R. N. Lyman, T. W. Kornack, and M. V. Romalis, Phys. Rev. Lett. \textbf{89}, 130801 (2002).

\bibitem{Kominis2003} I. K. Kominis, T. W. Kornack, J. C. Allred, and M. V. Romalis, Nature \textbf{422}, 596 (2003).


\bibitem{Smullin2006} S. J. Smullin, I. M. Savukov, G. Vasilakis, R. K. Ghosh, and M. V. Romalis, Phys. Rev. A \textbf{80}, 033420 (2009).

\bibitem{Savukov2005a} I. M. Savukov, S. J. Seltzer, M. V. Romalis, and K. L. Sauer, Phys. Rev. Lett. \textbf{95}, 063004 (2005).

\bibitem{Lee2006} S.-K. Lee, K. L. Sauer, S. J. Seltzer, O. Alem, and M. V. Romalis, Appl. Phys. Lett. \textbf{89}, 214106 (2006).



\bibitem{Seltzer2007} S. J. Seltzer, P. J. Meares, and M. V. Romalis, Phys. Rev. A \textbf{75}, 051407(R) (2007).

\bibitem{Kitching2002} J. Kitching, S. Knappe, and L. Hollberg, Appl. Phys. Lett. \textbf{81}, 553 (2002).

\bibitem{Cates1988} G. D. Cates, S. R. Schaefer, and W. Happer, Phys. Rev. A \textbf{37}, 2877 (1988).

\bibitem{Corney1977} A. Corney, \textit{Atomic \& Laser Spectroscopy} (Clarendon, Oxford, 1977).

\bibitem{Knappe2006} S. Knappe \textit{et al.}, J. Opt. A \textbf{8}, S318 (2006).

\bibitem{Lwin1978} N. Lwin and D. G. McCartan, J. Phys. B \textbf{11}, 3841 (1978).

\bibitem{Andalkar2002} A. Andalkar and R. B. Warrington, Phys. Rev. A \textbf{65}, 032708 (2002).

\bibitem{Li2006} Z. Li, R. T. Wakai, and T. G. Walker, Appl. Phys. Lett. \textbf{89}, 134105 (2006).

\bibitem{Rosenberry2007} M. A. Rosenberry, J. P. Reyes, D. Tupa, and T. J. Gay, Phys. Rev. A \textbf{75}, 023401 (2007).

\bibitem{McGillis1967} D. A. McGillis and L. Krause, Phys. Rev. \textbf{153}, 44 (1967).

\bibitem{McGillis1968} D. A. McGillis and L. Krause, Canadian J. Phys. \textbf{46}, 25 (1968).

\bibitem{Hrycyshyn1970} E. S. Hrycyshyn and L. Krause, Canadian J. Phys, \textbf{48}, 2761 (1970).

\bibitem{Rosen1999} M. S. Rosen, T. E. Chupp, K. P. Coulter, R. C. Welsh, and S. D. Swanson, Rev. Sci. Instr. \textbf{70}, 1546 (1999).

\bibitem{Bear2000} D. C. Bear, Ph.D. thesis, Harvard University (2000).


\bibitem{Masnou1967} F. Masnou-Seeuws and M. A. Bouchiat, Journal de Physique \textbf{28}, 406 (1967).

\bibitem{Silver1984} J. A. Silver, J. Chem. Phys. \textbf{81}, 5125 (1984).

\bibitem{Franz1976} F. A. Franz and C. Volk, Phys. Rev. A \textbf{14}, 1711 (1976).

\bibitem{Fedchak1997} J. A. Fedchak, P. Cabauy, W. J. Cummings, C. E. Jones, and R. S. Kowalczyk, Nucl. Instr. Meth. Phys. Res. A \textbf{391}, 405 (1997).

\bibitem{Happer1977} W. Happer and A. C. Tam, Phys. Rev. A \textbf{16}, 1877 (1977).

\bibitem{Appelt1998} S. Appelt, A. Ben-Amar Baranga, C. J. Erickson, M. V. Romalis, A. R. Young, and W. Happer, Phys. Rev. A \textbf{58}, 1412 (1998).

\bibitem{Happer1967} W. Happer and B. S. Mathur, Phys. Rev. \textbf{163}, 12 (1967).

\bibitem{Anderson1960} L. W. Anderson, F. M. Pipkin, and J. C. Baird, Phys. Rev. \textbf{116}, 87 (1959).

\end{thebibliography}
\end{document}